\documentclass{JFM-FLM_Au}

\usepackage{amsmath}
\usepackage{amsfonts}
\usepackage{float} 
\usepackage[toc,page]{appendix}
\usepackage{graphicx}
\usepackage{caption}
\usepackage{subcaption}
\usepackage[utf8]{inputenc}
\usepackage{booktabs}
\usepackage{units}

\usepackage{adjustbox} 
\usepackage{multirow} 

\usepackage{array}

\lefttitle{Etienne Boulais and Richard D. Braatz}
\righttitle{Journal of Fluid Mechanics}

\title{A ``Lego Block" Approach to Flow in Complex Microfluidic Networks}

\author{Etienne Boulais\aff{1},
 \and Richard D. Braatz\aff{1}}

\affiliation{\aff{1}Department of Chemical Engineering, Massachusetts Institute of Technology, 77~Massachusetts Avenue, Cambridge, MA 02139, USA}

\corresau{Richard D. Braatz, \email{braatz@mit.edu}}

\begin{document}

\maketitle

\begin{abstract}
We present a new way to construct analytical solutions for flow in complex microfluidic channel networks, as well as planar disordered media. Using a combination of Schwarz-Christoffel maps and segmentation techniques inspired by integrated circuit analysis, we build a library of base building blocks which can be reassembled to model complex geometries, in the style of ``Lego Blocks''.
Our approach requires minimal numerical computation, and can then generate analytical solutions for any combination of inlet and outlet flow rates. Moreover, our method can tackle multiply connected domains which are usually difficult to model using typical conformal transform approaches.
The solutions are developed for microfluidic Hele-Shaw cell devices, but also apply to ideal flow and Darcy flow in complex geometries, or any other flow problem adequately modeled by Laplace's equation. We end by showing how the procedure can be used to model complex disordered media, fractal-like flow geometries, as well as problems of steady advection-diffusion in microfluidic mixers.
\end{abstract}

\section{Introduction}



Flow in complex disordered media is an active topic of research in geophysics \citep{fagherazzi2012numerical, zhang2020global}, groundwater mechanics \citep{sahimi1993flow, scheibe1998scaling}, hydrocarbon recovery \citep{blunt2013pore}, chemical engineering \citep{coppens1995diffusion, nemec2005flow, mujeebu2009applications}, and more recently life sciences \citep{yamada2019mechanisms, henke2020extracellular}.
Classic approaches to linking microscopic, pore-scale flow and macroscopic mean-field behavior has included statistical continuum theory \citep{torquato2002random} and percolation theory \citep{sahimi1994applications}. 
From the 1980s, disordered complex media was modeled as networks of throats connected by channel-like pores \citep{chandler1982capillary, koplik1982creeping, toledo1994pore}. More recently, wider availability of numerical tools has enabled increasingly sophisticated studies of similar problems related to elastic turbulence \citep{browne2021elastic}, vortex formation in porous media \citep{residori2025flow}, as well as multiphase flow in complex microenvironments \citep{kang2014pore}, just to name a few. 
%
%
Interest in such studies has been renewed in the past decade with the increasing availability of microfluidic model systems \citep{anbari2018microfluidic, zhao2016wettability, strom2024geometry}, which give scientists a new and unprecendeted experimental access to microscopic features of these flows in a very controlled environment.
This movement is part of a wider contemporary trust in modern physics on the study of the physics of complex systems (see for example \cite{restagno2024prospectives}), where the combination of modern numerical tools, analytical methods, and experimental technologies enable a broader understanding of multiscale problems, allowing scientists to start bridging microscopic pore-scale descriptions with macroscopic mean-field behavior in microstructured media.


One major difficulty in developing theoretical models for flow in disordered media is the inherently irregular geometry which these flows exist in. A tool of choice for analyzing flow in complex geometries is Schwarz-Christoffel maps. 
Part of a larger family of conformal maps, a classic tool in the analysis of ideal flow \citep{milne1996theoretical} and flow in porous media \citep{strack1988groundwater}, Schwarz-Christoffel map are used to transform solutions of Laplace equations in canonical domains to complex polygonal geometries.
Early applications of these maps were limited by the lack of tools for efficiently computing maps to large polygons, with advanced analytical work mostly centered around symmetrical or otherwise simplified domains \citep{seeger1953diffusion, versnel1979analysis, he2014calculating}.
Computation of Schwarz-Christoffel maps has been made more broadly available by efficient methods for its numerical computation, developed in the 1980s \citep{trefethen1980numerical}. They have since been used in fluid mechanics for the generation of meshes in complex channel geometries \citep{floryan1985conformal}, in studying free surface flow colliding with polygonal obstacles \citep{elcrat1986classical}, and more recently as a tool for modeling transport in microfluidic chambers \citep{boulais2025steady}.
Schwarz-Christoffel maps do, however, remain an underused tool in fluid physics, perhaps due to their perceived complexity, as well as the ease of use and wide availability of simpler numerical tools such as finite difference or finite element methods. Used wisely, however, they can enable very sophisticated analytical work that is not possible to do using typical numerical methods.

In this paper, we model flow in complex media and large-scale microfluidic circuits using Schwarz-Christoffel maps.
To do so, we take inspiration from classic work on the analysis of integrated circuits \citep{horowitz2004resistance}.
By combining efficient numerical computation of the Schwarz-Christoffel maps and previously published solutions for multipolar flow in Hele-Shaw cells \citep{boulais2020two}, we construct a library of base elements (or ``lego blocks'') which can be recombined to construct elaborate flow geometries. As such, our approach is bottom-up rather than top-down. Once a library of building blocks has been constructed, its elements can be recombined at will at little to no additional computing cost, allowing us to build solutions for flow in arbitrarily complex geometries.
The solutions we build have direct applications in the study of microfluidic large-scale integration \citep{thorsen2002microfluidic}, microfluidic models for porous media \citep{anbari2018microfluidic}, as well as problems of groundwater flow in complex aquifers \citep{haitjema1995analytic}. We also show how the solutions can be directly extended to include the study of diffusion, making them also ideal for modeling complex crystallization processes in microfluidic chips \citep{li2010protein, shi2017progress,  belliveau2012microfluidic}.

%
The work presented therein is an extension of our previously published work on analysis of polygonal microfluidic channel junctions \citep{boulais2025steady}. We extend some of the ideas included in the paper (in particular the ``stitching together" approach to solution construction) and show how it can be extended to the study of very complex flow geometries, at virtually no additional computing cost.
The work presented here is uniquely enabled by the Schwarz-Christoffel map and analytical solutions for multipolar flow, and allows the flexible analysis of problems which would be difficult to treat using more direct numerical approaches such as finite difference or finite element methods. By constructing and combining basic geometric building blocks, we are also bypassing a lot of the usual constraints which come with flow modeling using conformal maps, namely with regards to the connectivity of the domain. In particular, our method allows the analysis of multiply connected domains naturally, without requiring additional sophistication in the calculations.
The main innovation of the present work is in the bottom-up construction of flow solutions, in our method for constructing analytical flow maps by trimming multipolar solutions, as well as in the wide range of flow geometries we can tackle.








\section{Outline}

\begin{figure}[h]
	\centering
    \includegraphics[width=0.95\textwidth]{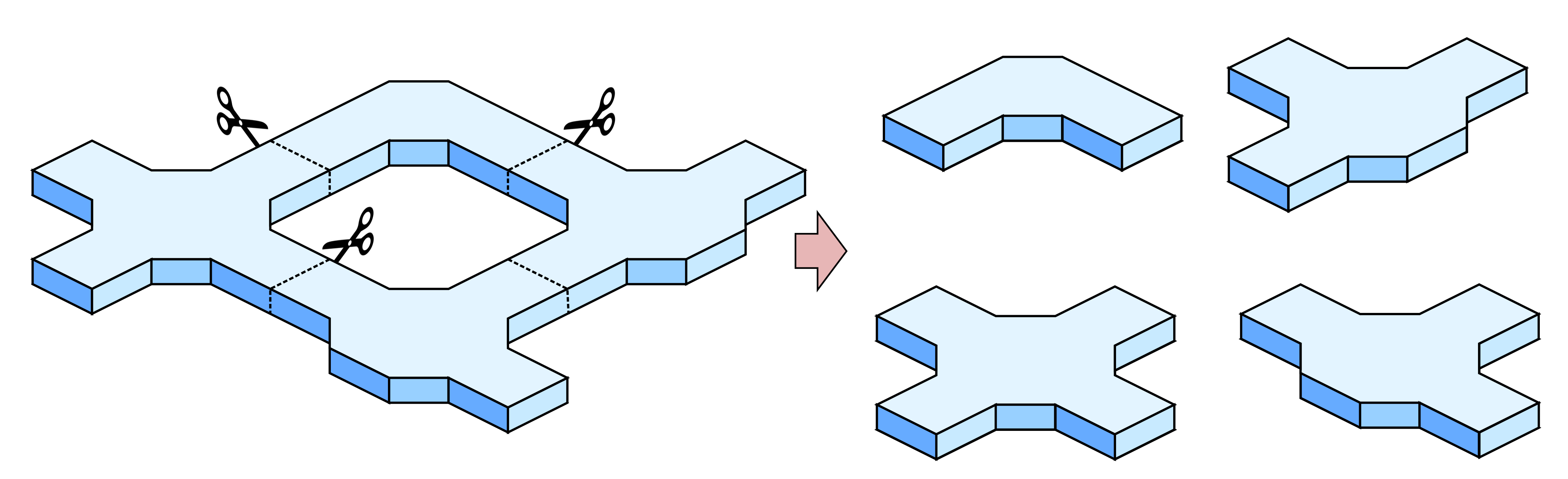}

	\caption{Schematic representation of our approach. A complex microfluidic circuit is subdivided into simpler individual junction elements, which can then be analyzed individually and recombined. Conversely, a library of smaller elements can serve as building blocks for complex circuits.}
	\label{fig_problem_schematic}
\end{figure}

Before beginning, it is worth outlining the overall strategy we use throughout this article for building analytical flow solutions. The details of the procedure will be specified in further sections.
We are searching for solutions for flow in complex microfluidic systems such as the one illustrated in Fig. \ref{fig_problem_schematic}. If the device is shallower than the characteristic in-plane dimension, and flow rates are sufficiently low, it is appropriate to use a description in terms of a Hele-Shaw cell \citep{boulais20232d}, which reduces the problem to that of solving the Laplace equation in a complex geometry. Taking inspiration from work on integrated circuit analysis \citep{horowitz2004resistance}, we subdivide our complex geometry into a number of smaller polygonal ``blocks'', which can then be stitched together.
Each individual block can be modelled as a Laplace boundary-value problem in a polygonal geometry, with no-flux conditions prescribed on walls and fixed potential value on the inlets / outlets.
Using Schwarz-Christoffel transforms, we reduce this boundary-value problem in a polygon to a boundary-value problem on a disk, which can then be further transformed into an arrangement of point sources and sinks in the half-plane, for which analytical solutions are easily formulated.
Once this has been done for every block, they can then be reassembled, and the exact value of the potential at each inlet / outlet can be obtained by using classic tools of circuit analysis (see for example \cite{bruus2007theoretical}).

The main advantage of this method is that the ``hard'' numerical step, the conformal map from the polygonal block to the disk, only has to be computed once for each block geometry. After that, solutions can be generated virtually instantly for any combination of flow parameter within this block. Furthermore, from even just a handful of basic building blocks, one can then proceed to construct flow circuits of arbitrary complexity.
Beyond just obtaining the flow profile everywhere in our system, the solutions can be directly extended to include other physical phenomena, such as steady advection-diffusion, at little to no additional cost, as is done in section \ref{sec_diffusion}.

\section{Theory}

We are building solutions for microfluidic flows in complex geometries. Because of their fabrication, many microfluidic systems can be described as Hele-Shaw cells, or quasi-2D domains confined between two parallel plates, where the gap between the plates is smaller than the characteristic in-plane dimension \citep{boulais20232d}.
The velocity profile in a Hele-Shaw cell (that is, a thin domain with a small constant thickness $d$ in the $z$ direction) is given by \citep{batchelor1967introduction}

\begin{equation}
	\vec{u} \left( x, y, z \right) = - \frac{1}{2 \mu} z \left( d - z \right) \left( \frac{\partial p}{\partial x} \vec{x} + \frac{\partial p}{\partial y} \vec{y} \right)
\end{equation}

Where $\mu$ is the fluid viscosity and $p$ is the hydrostatic pressure. This is equivalent to the gradient of a 2D potential function, modulated by a parabolic flow profile in the out-of-plane direction. The potential function $\phi \left( x, y \right)$, which is proportional to $p \left( x, y \right)$, respects Laplace's equation

\begin{equation}\label{eq_laplace}
    \nabla ^2 \phi = 0
\end{equation}

This description is valid everywhere except in a small boundary layer near the channel walls, whose thickness scales with the ratio of the channel thickness to the in-plane characteristic dimension \citep{gokcce2019self}.
The problem of describing flow within a complex microfluidic system can thus be reduced to that of solving Laplace's equation, with no-flux boundary conditions at the channel walls, as well as appropriate Dirichlet or Neumann conditions at the inlets and outlets, depending on whether we want to prescribe pressure of flux at each aperture.
%
%
Outside of simple geometries, such problems are usually solved numerically using finite difference, finite volume, or finite element methods. However, direct numerical approaches can be slow, difficult to scale as the systems get bigger, and require a new computation for any new combination of input parameters (inlet flow rates, flow rate ratio, etc).
%
%
What we do instead is subdivide our complex fluidic circuit into a number of individual elements along ``break lines", which are chosen so as not to disturb the flow. Break lines are chosen so that the potential function is approximately constant along them (or conversely so that the streamlines intersect the break line at approximately right angles). This is similar to classic approaches to the modeling of electric fields in integrated circuits \citep{horowitz2004resistance}. 
This subdivision reduces the problem of finding the potential in a fluidic circuit to a series of smaller problems in individual circuit elements (as well as the problem of their subsequent reassembly).
Conversely, our method allows one to build a library of individual circuit elements which can be used as building blocks to create circuits of arbitrary complexity at little cost. 
We now proceed to show how these individual elements can be analyzed.

%



\section{Conformal mapping}


Description of our flow problem in terms of the 2D Laplace's equation allows us to use conformal mapping as part of our solution procedure. Coordinates in 2D space can be written as a single complex number $z = x + i y$, and flow can be described using the complex potential $\Phi = \phi + i \psi$, whose real part $\phi$ corresponds to the level set of the classic potential function while the imaginary part $\psi$ is the streamfunction \citep{milne1996theoretical}. 
An immense library of existing conformal maps \citep{kythe2019handbook} can be used to transform problems in complex geometries to more tractable ones. Specifically, conformal maps can be used to reduce problems in seemingly different geometries to a single ``canonical" geometry where they are more easily solved.
We are interested in circuits with complex-shaped elements, so we use the Schwarz-Christoffel map \citep{driscoll2002schwarz}, which maps between a disk domain $z$ and an arbitrary polygonal domain $w$:

\begin{equation}\label{eq_schwarz_christoffel}
	w \left( z \right) = w_c + C \int^z \prod_{k=1}^{n} \left( 1 - \frac{\xi}{z_k} \right)^{\alpha_k - 1} d \xi  
\end{equation}

Where $w_c$ and $C$ are parameters corresponding to translation and scaling of the map, and the $\alpha_k$ correspond to the interior angles of the polygon, in units of $\pi$.
The challenge in computing this map is that it has a set of unknown parameters $z_k$, which correspond to the image of the polygon's vertices on the circumference of the circle in the $z$ domain. For polygons with 3 or less vertices, the Riemann mapping theorem allows us to pick these prevertices freely, but for larger polygons, they have to be determined as part of the solution procedure. No simple analytical method exists for solving this ``parameter problem" of the Schwarz-Christoffel map, but very efficient numerical methods are available \citep{trefethen1980numerical}.
%
Fortunately, the parameter problem only has to be solved once for any given element geometry.
%
We have shown in a previous publication \citep{boulais2025steady} how Schwarz-Christoffel maps could be used to model diffusion in polygonal microfluidic junctions. Here we extend this work and show how elements built using the same method can be used as building blocks for complex fluidic circuits. 

Conformal mapping approaches, in particular those using the Schwarz-Christoffel map, can have trouble with a number of pathological domains. Domains with elongated regions can rapidly become difficult to treat, as the condition of conformality forces large regions of the polygonal domain to be mapped to vanishingly small regions of the disk. Domains containing very narrow ``choke points" can become difficult to treat numerically for the same reasons, although sophisticated methods do exist to treat those problems \citep{driscoll1998numerical}. 
The extension of conformal mapping to multiply connected domains is also a non-trivial problem, with solutions existing for only a few families of carefully constructed geometries \citep{embree1999green}. 
Our approach gets rid of some of these difficulties. Complex microfluidic circuits which include elongated segments, large variations in aspect ratios, and several ``holes" in the domain, can usually be subdivided to a small number of individual elements which each have manageable aspect ratios and are simply connected.

\section{Problem in the Disk Domain}

\begin{figure}[h]
	\centering
	\includegraphics[width=0.9\textwidth]{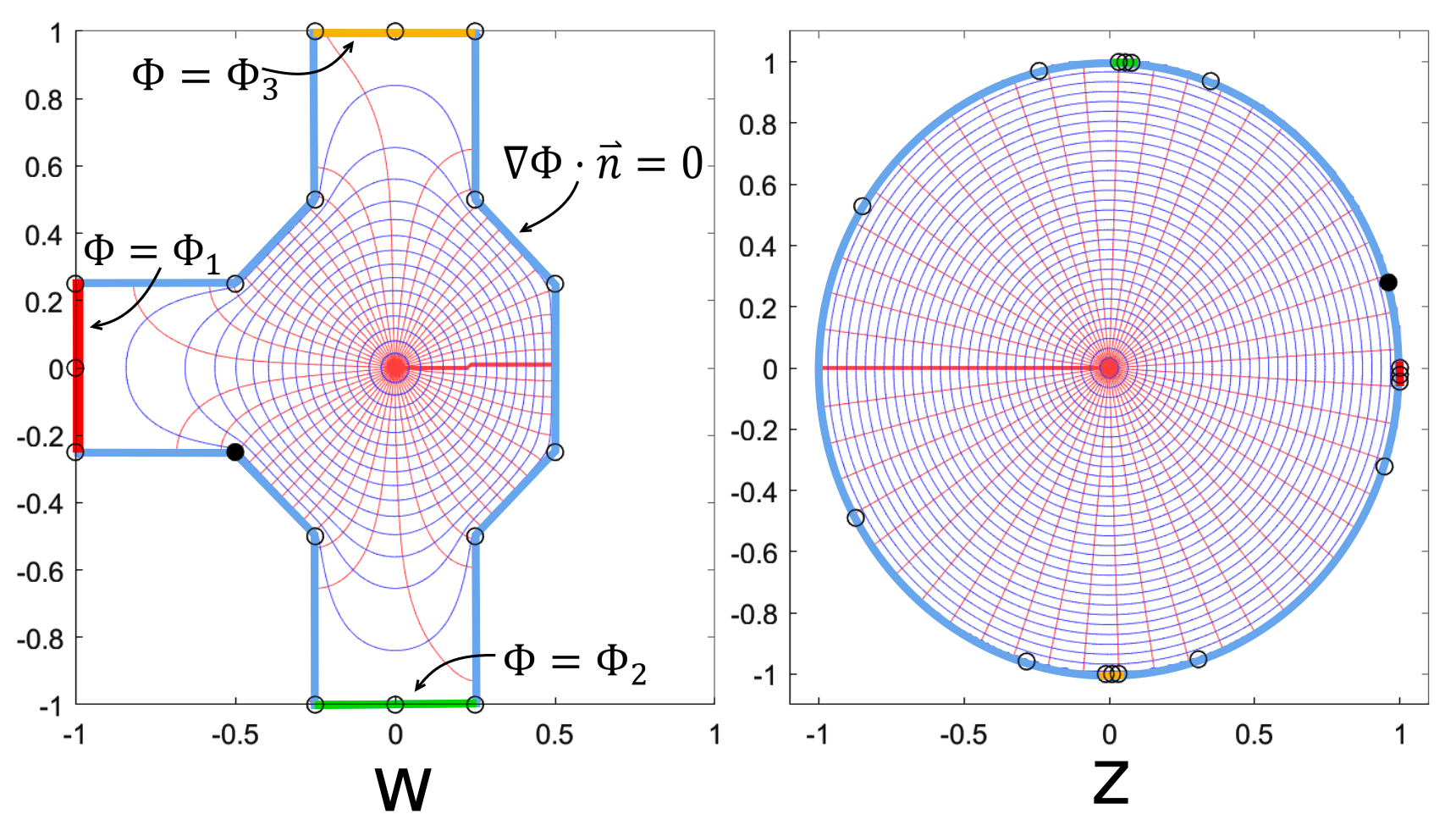}

	\caption{Transforming the boundary value problem in the channel junction element (w domain) to an equivalent problem in a disk domain (z domain). No-flux conditions are prescribed on the walls, and fixed value of the real potential are set on each inlet and outlet.}
	\label{fig_disk_domain}
\end{figure}

Application of the Schwarz-Christoffel map \ref{eq_schwarz_christoffel} allows us to simplify the boundary value problem for a single polygonal element to one in the interior of the unit disk, as illustrated in Figure \ref{fig_disk_domain}.
%
The main challenge is that the resulting problem in the interior of the disk is a mixed boundary-value problem. It has prescribed value of the potential function (Dirichlet boundary conditions) on circular arc segments which correspond to the connecting edges of an element, as well as zero-flux Neumann conditions on the rest of the circle's perimeter. The exact position of each segment will depend on the position of the prevertices in the disk domain.
%
An analytical solution of this mixed boundary value problem is in the general case very difficult, and requires subtle semi-analytical approaches \citep{wendland1979integral}. 
Conversely, while numerically solving the boundary value problem for a set of prescribed potential values at each port would be relatively straightforward, using either finite difference formulas or simple relaxation methods, this would be impractical as it would require a new numerical step for any combination of port potentials. 
In the following sections, we show how this difficult mixed boundary value problem can be circumvented, and simple analytical solutions for the potential can be found. 
This allows us to limit the numerical work to the computation of the Schwarz-Christoffel map and its inverse for each geometrical element in our library. We can then build circuits of arbitrary complexity from our small library of previously defined elements.

\subsection{Flow within a single block}

Flow within a 2-port element (such as a channel segment) only has to be solved for one set of potentials, and the solution for any other set can be trivially obtained by rescaling and addition of a constant factor.
Any 2-port resistive element can be described by a single scalar, called its resistance, linking the total flow through the element to the difference in potential at its edges \citep{bruus2007theoretical}. Schwarz-Christoffel maps have been used in the past for computing the resistance of plane circuit elements, by mapping complex shapes to uniquely determined rectangles \citep{trefethen1984analysis}. 
In the case of 2-port elements, conformal mapping does offer some advantages over typical numerical approaches (for instance in the treatment of sharp corners), but the main advantage of our method is its extension to multi-port elements.


The situation becomes considerably more complex as we start considering elements with more than 2 ports. 
While the complex potential in a 2-port element is, up to rescaling and addition of a scalar, a unique function of the element's geometry, potential in a 3-port element becomes a function of the geometry as well as an additional scalar expressing flow rate ratio or the ratio of the fixed real potential between two of the inlets. Likewise, flow in a 4-port element will depend on the geometry as well as 2 scalars, etc. 
%
Methods have been proposed in the past to model the Laplace equation in junctions using Schwarz-Christoffel maps. Modi et al. \citep{modi1981conformal} used conformal maps to find the location of stagnation points in specific geometries of channel junctions. Their work was extended by Best et al. \citep{best1984separation} who used a similar method to determine the separation regions in similar channel junctions at high Reynolds numbers.
More recently, in studying electrostatics problems, Wang et al. \citep{wang2016conformal} proposed a way to construct equivalent representations for multi-port elements in which the potential takes on a simple form. In their work, however, a different Schwarz-Christoffel map has to be computed for each combination of prescribed potential at the ports. 
%
%
We propose an approximate way of solving the problem which, once the Schwarz-Christoffel map has been computed, makes the flow within the mixer expressible as a simple analytical expression. Our approach differs from previously mentioned sources (such as \cite{wang2016conformal}) in that a single map has to be computed for a given element geometry, regardless of the potential at the ports. This is, to the best of our knowledge, an innovation of this paper. 

\begin{figure}[h]
	\centering
    \includegraphics[width=0.95\textwidth]{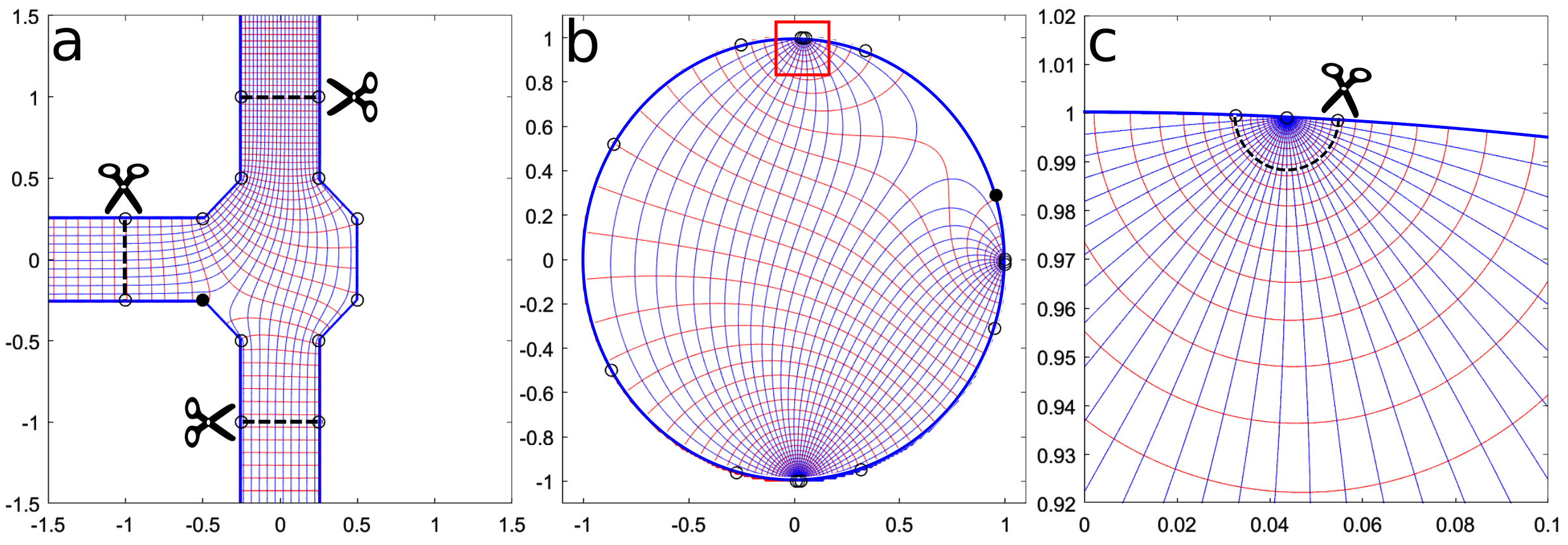}

	\caption{Truncating a channel junction with outlets extending to infinity to create a finite piece. a. Illustration of a junction and the cut lines. b. The same flow problem in the disk domain. c. Zoomed-in version of the flow domain (red square in subfigure b) showing the image of the cut line}
	\label{fig_truncation}
\end{figure}

Instead of finding a map to the interior of a bounded polygon representing our element, we compute a map to an equivalent polygon with ports extending to infinity (Fig. \ref{fig_truncation}a). This allows us to place a single point source of known strength for the image of each inlet in the disk domain. Using the Moebius map

\begin{equation}\label{eq_moebius_transform}
	\nu = i \frac{z + z_0}{z - z_0}, 
\end{equation}

we can map the disk domain $z$ to the upper half-plane $\nu$, in which the complex potential is expressed as a simple sum of logarithmic terms

\begin{equation}\label{eq_sumoflogs}
	\Phi ( \nu ) = \sum_i k_i \text{ln} ( \nu - \nu_i )
\end{equation}

Where the sum is over all inlets and outlets, $k_i$ is the dimensionless flow rate at each inlet and outlet, and $\nu_i$ is the image of the vertex corresponding to that inlet in the $\nu$ domain. 
Once a solution for the unbounded mixer has been obtained, it can then be cropped to keep only the bounded polygonal section we need (see Fig. \ref{fig_truncation}). The conditions for such a cropping to be acceptable are the same as the ones required to subdivide our complex fluidic network into a series of individual elements, which is that the potential function has to be approximately constant along the cut line (which means that the cut has to happen far enough from turning regions in the flow).
%
Equation \ref{eq_sumoflogs} describes the flow in terms of the flow rates at the ports, and not the value of the potential at the cut lines. In a simple 2-port resistor, the flow through the element and the difference in potential are linked by the resistance using Ohm's law, or its fluidic analog

\begin{equation}
    \Delta \Phi = R Q
\end{equation}

In multi-port elements, we can easily go from a representation in terms of flow rates to one in terms of potentials by evaluating equation \ref{eq_sumoflogs} for a set of flow rates, and directly extracting the potential at each edge. To do the reverse, and obtain the flow rates $k_i$ from prescribed values of the potential at each edge, we use a representation of the element in terms of a network of resistor (see Appendix \ref{appendix_resistors}).


\section{Assembling Elements}\label{sec_assembling}

\begin{figure}[h]
	\centering
    \includegraphics[width=0.95\textwidth]{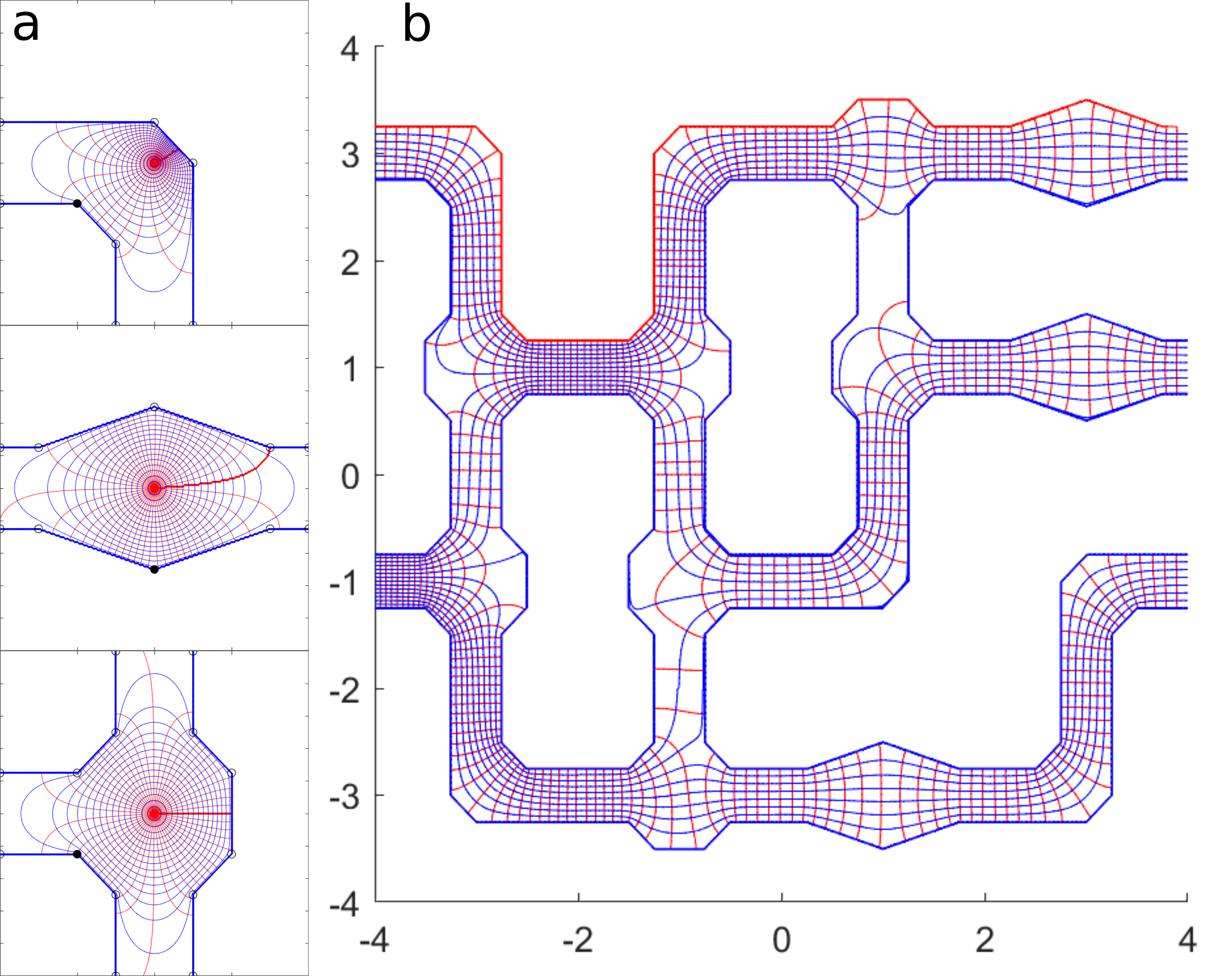}

	\caption{Assembling unit elements to generate flow in a multiply connected domain. a. Schwarz-Christoffel maps precomputed for three individual building blocks. b. Assembling the three unit blocks to generate a complex circuit. Value of the potential at each element junction was determined by solving the equivalent resistive circuit problem. Blue lines are streamlines while red lines are the level sets of the real potential.}
	\label{fig_assembly}
\end{figure}

Once the Schwarz-Christoffel map to individual building blocks has been computed, and their equivalent resistance (or network of resistances) has been determined, they can be assembled to form complex circuits, as is shown in Figure \ref{fig_assembly}. Flow through each end of each element, as well as the value of the potential function at the extremities, are computed by solving the equivalent resistive circuit. This can be done by explicitly writing down Kirchoff's laws for smaller circuit, or using freely available linear circuit solvers for larger circuits.
%
The complex potential at each element is also offset by an imaginary constant in order to ensure that the streamfunction matches correctly at each break line. In the figures in this article, we have arbitrarily fixed the streamfunction to be 0 at the top-left corner of our system, with every other element being offset so that the streamfunction is continuous at the break lines.
In this article, we have only been modeling planar circuits: circuits which can be plotted on a flat plane without having overlapping or intersecting elements. If we allow different channel segment to intersect one another, we may generate pathological geometries where it is impossible to correctly match the streamfunction at every break line.

\section{Models for Porous Media}\label{sec_porous}

\begin{figure}[h]
	\centering
    \includegraphics[width=0.95\textwidth]{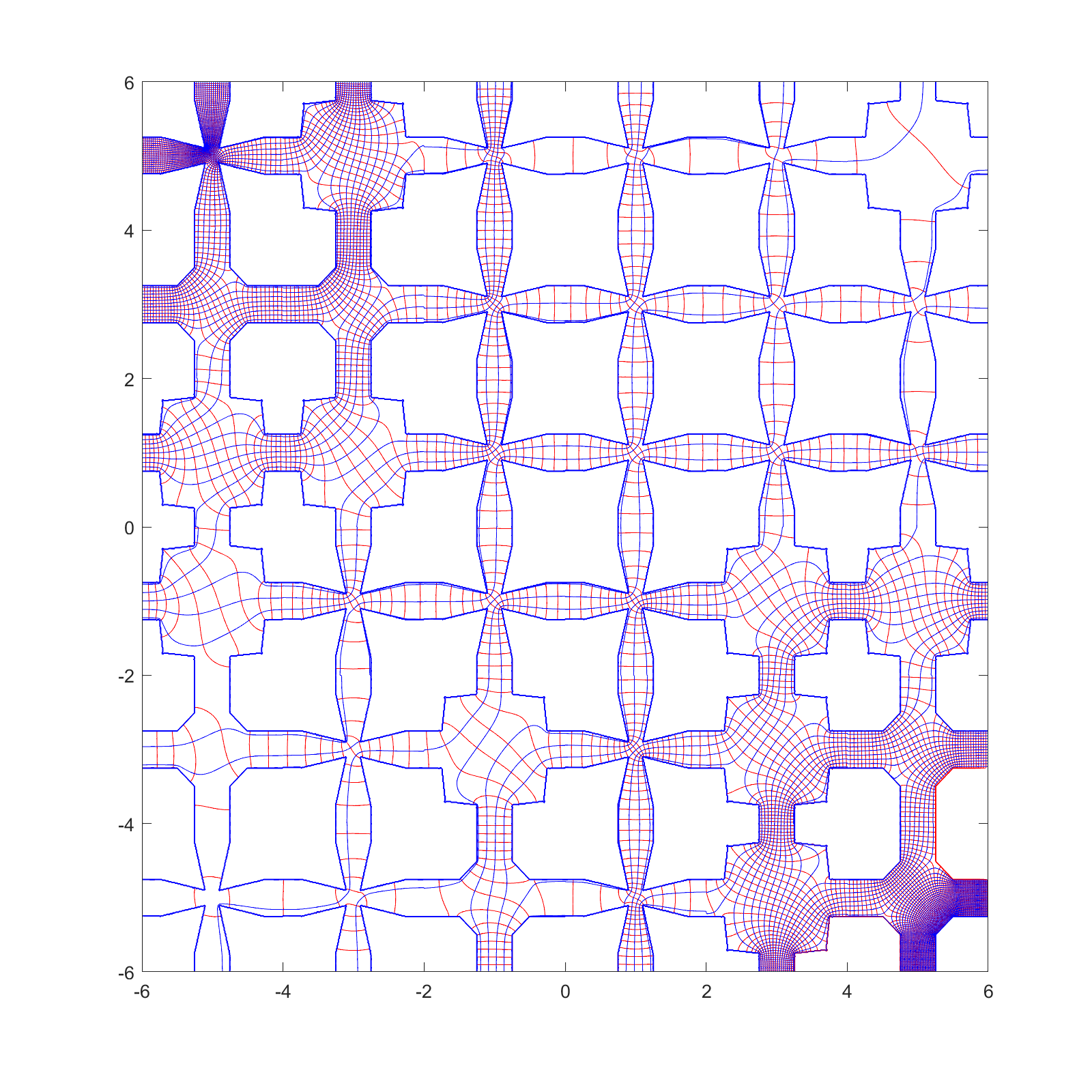}

	\caption{Model porous media obtained by assembling junction unit elements}
	\label{fig_porous}
\end{figure}

One application of the method presented here is in the description of flow in microfluidic models of porous media.
There has been a renewal of interest in recent years on microscopic models of porous media, fueled in part by advances in microfluidics, which enable the fabrication and imaging of precise micromodels of flow in fractured rocks, soils, and other disordered media \citep{anbari2018microfluidic}.
Connections between microscopic, pore-scale description of flow and continuous effective medium has a long history in the analysis of heterogeneous media \citep{torquato2002random}.
Our representation of individual elements in terms of arrangement of resistors (Appendix \ref{appendix_resistors}) hints at classic work on percolation in random networks of resistors \citep{straley1977critical}. This type of modeling has found applications in soil physics \citep{berkowitz1998percolation} and the study of flow through fractured rocks \citep{viswanathan2022fluid}. 
Regarding explicit analysis of flow in such systems, in the 1980s, Koplik analyzed Stokes flow in random networks constructed from a combination of circular "throats" interconnected by channel-like "pores" \citep{koplik1982creeping}, matching pressure and velocity between adjacent elements. Combination of polygonal junction elements, modeled using Monte-Carlo simulations, were also used to model porosimetry in rock samples
\citep{toledo1994pore}.
%
From a handful of possible channel junction geometries, we build a model porous media, illustrated in Fig. \ref{fig_porous}. Values of the complex potential are fixed at the edges of the domain, and the potential everywhere within the network is obtained using a simple relaxation method. More discussion on the analysis of large networks of random elements can be found in classic reviews on percolation and conduction \citep{kirkpatrick1973percolation}.
Most previous microscopic descriptions of flow in porous media focuses either on throats, with simplified channels connecting chamber domains, or on pore channels connected by simple nodes (see for example \cite{koplik1982creeping} or \cite{chandler1982capillary}). Our description does not require a clear distinction between "pores" and "throats", and thus can enable the study of more general flow geometries which aren't covered by these previous approaches.

Beyond analysis of porous media, this construction could be very useful in modeling geometry-dependent flows in regular arrays of micropillars in microfluidic systems \citep{strom2024geometry, browne2020pore}.
Significant work already exists in percolation theory and adjacent fields for bridging network representations to effective scalar properties of a medium \citep{sahimi1994applications}. Here we provide a bridge between the more abstract network description and a microscopic description of flow, in a framework that allows for a wide range of possible geometries.
Complete description of the flow within these systems also enables the study of more complex particulate movement within them. Flow within such model porous media could be used as a starting point for more sophisticated studies of colloid flocking \citep{morin2017distortion}, anomalous diffusion \citep{bouchaud1990anomalous}, or transport of motile particles \citep{kurzthaler2021geometric} in disordered systems, just to name a few.

\section{High Aspect Ratio Domains}

\begin{figure}[h]
	\centering
	\begin{subfigure}{0.45\textwidth}
		\includegraphics[width=\textwidth]{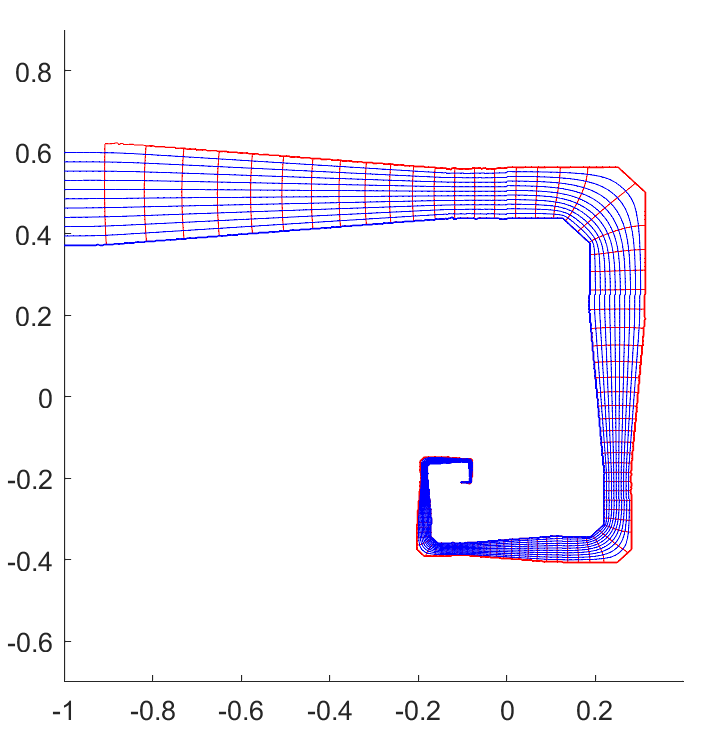}
	\end{subfigure}
	\begin{subfigure}{0.45\textwidth}
		\includegraphics[width=\textwidth]{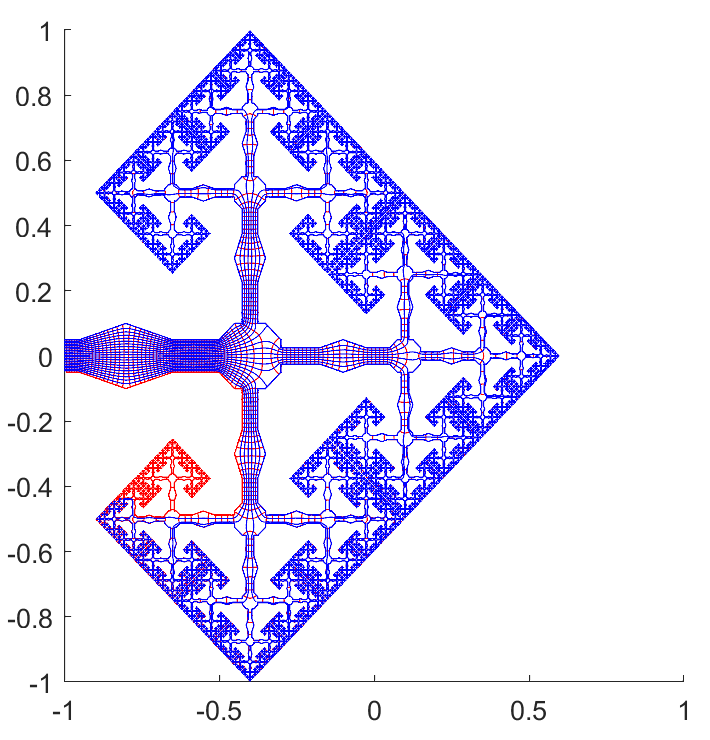}
	\end{subfigure}
	
	\caption{High aspect ratio domains. a. A spiral channel built by combining a shrinking channel segment and a corner piece. b. A fractal tree-like structure obtained by connecting scaled versions of a single junction element.}
	\label{fig_high_aspect_ratio}
\end{figure}

%
Our models can be easily applied to the study of domains with highly varying aspect ratios. 
Different pieces can be built which have different sized inlets and outlets. Moreover, the same basic building block can easily be scaled up or down in order to fit on a connecting outlet. This allows us to create domains with highly varying aspect ratio, as illustrated in Fig. \ref{fig_high_aspect_ratio}. In Fig. \ref{fig_high_aspect_ratio}a, we show how a shrinking spiral channel can be built from a corner piece as well as a ``shrinking channel'' segment. In Fig. \ref{fig_high_aspect_ratio}b, we build a fractal network of pores and throats by connecting identical versions of a channel junction with shrinking outlet elements. These sorts of geometries are no more complicated to build than the other channel networks shown in this article. Typical Schwarz-Christoffel based approaches often struggle when directly tackling high aspect ratio domains, due to the phenomenon of ``crowding'' imposed by the conformality condition \citep{driscoll2002schwarz}, and while sophisticated methods do exist to bypass these problems \citep{driscoll1998numerical}, they can be quite complex to implement and do not always mesh with every possible geometry.
The method presented here, by contrast, can treat these pathological domains quite naturally. The availability of complete flow maps for chambers with high aspect ratio variations as well as fractal channel networks opens up exciting possibilities for the modeling of flow in complex geometries.
In particular, this could enable further study of flow in the backbone of percolation clusters, which are known to have highly varying aspect ratio and fractal geometry \citep{herrmann1984building}. Flow in such fractal geometry has direct applications in the study of hydrology \citep{berkowitz1993percolation}, and could also be applied to the study of catalysts \citep{coppens1995diffusion}, networks of cracks \citep{bouchaud1993models, bourrianne2021crack}, root systems of plants \citep{gerwitz1974empirical} or fungi \citep{oyarte2025travelling}, or any other number of flow network with fractal or fractal-like geometry.

\section{Integrating Diffusion}\label{sec_diffusion}

\begin{figure}[h]
	\centering
    \includegraphics[width=0.95\textwidth]{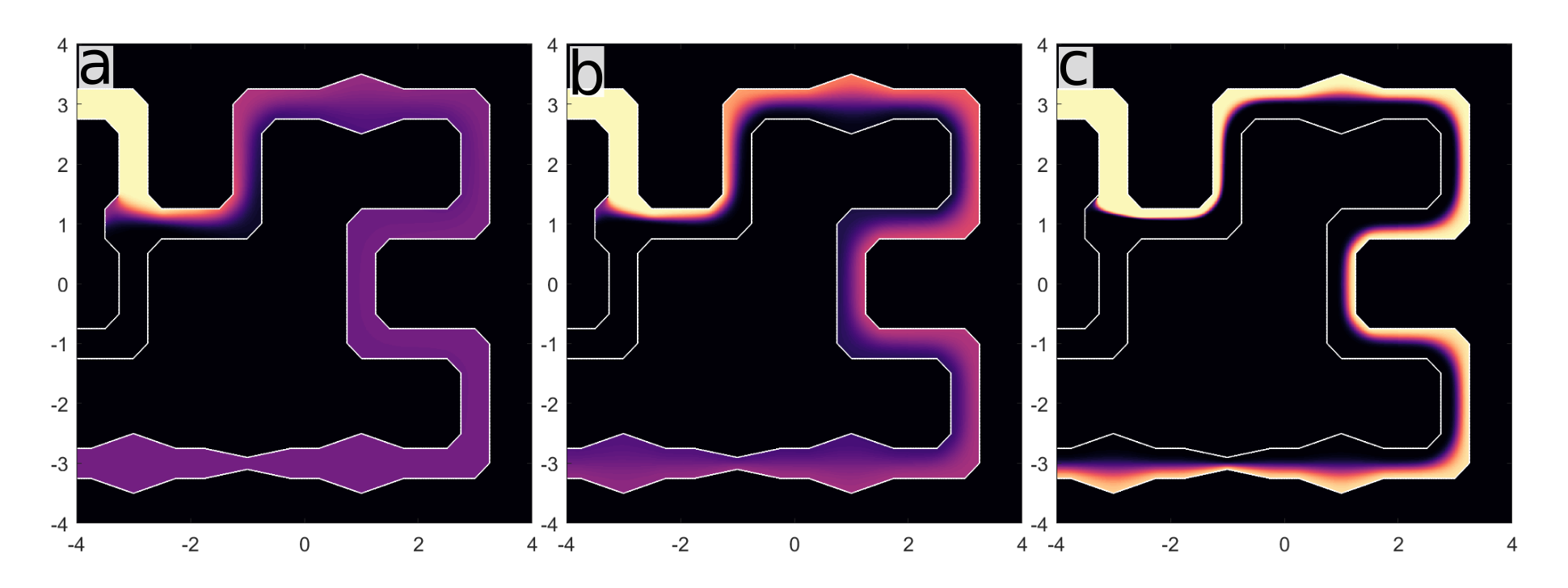}

	\caption{Steady advection-diffusion in complex microfluidic networks at different Peclet numbers. a. $\Pen = 50$. b. $\Pen = 200$. c. $\Pen = 2000$}
	\label{fig_diffusion}
\end{figure}

Formulation of our problem in terms of 2D potential flows means our model we can easily be extended to incorporate steady advection-diffusion. This could be used to track concentration profiles or temperature variation in the flows described above. 
Steady advection-diffusion in 2D potential flows is described by the equation
\begin{equation}\label{eq_advection_diffusion}
    0 = \nabla^2 c - \Pen \ \vec{u} \cdot \nabla c
\end{equation}
For many steady-state microfluidic systems, this description in terms of height-averaged concentration is adequate, see for example \cite{goyette2019microfluidic, boulais2025steady} for experimental validation and further discussion. 
Equation \ref{eq_advection_diffusion}, combined with the Laplace equation describing the potential flow, forms a conformally invariant system of PDEs \citep{bazant2004conformal}. We can readily use the same conformal mapping strategy to solve the associated diffusion problem. Specifically, the complex potential $\Phi \left( z \right) = \phi + i \psi$ maps the problem to streamline coordinates, where the advection-diffusion equation takes the simple form
\begin{equation}
    \frac{\partial^2 c}{\partial \phi^2} + \frac{\partial^2 c}{\partial \psi^2} = \Pen \frac{\partial c}{\partial \phi}
\end{equation}
The boundary conditions map to horizontal segments in the streamline domain. In the general case, this may lead to complex mixed boundary condition problems which can become quite laborious to solve (see for instance \cite{mckee2025steady} for approaches to this problem). We have previously shown how simple solutions can be developed for diffusion in polygonal microfluidic junctions \citep{boulais2025steady}, which we extend here to our complex constructed domains.
Using the method shown in section \ref{sec_assembling}, we construct a fluidic network consisting of a T-mixer \citep{kamholz1999quantitative} followed by an irregular serpentine channel. The boundary conditions for the diffusion problem are prescribed concentration at infinity in either inlet, as well as no-flux (zero derivative) conditions on every wall.
In the streamline domain, this problem maps to two strips of initially different concentrations which are initially separated by an impermeable boundary, then brought into contact at a point which corresponds to the image of the flow's stagnation point.
In the ``near-field", a simple solution to this problem can be constructed from error functions, as we have shown in previous publications \citep{boulais2020two}
\begin{equation}
	\label{solution_wake}
	c \left( \Phi \right) =
	\begin{cases} 
		\frac{1}{2} \left( 1 - \text{erf} \left( \text{Im} \sqrt{ \Pen \left( \Phi - \Phi_\mathrm{stag} \right) } \right) \right), & \psi < 0 \\
		\frac{1}{2} \left( 1 + \text{erf} \left( \text{Im} \sqrt{ \Pen \left( \Phi - \Phi_\mathrm{stag} \right) } \right) \right), & \psi \geq 0 \\
	\end{cases}
\end{equation}
Where $\psi$ is taken to be 0 on the separating streamline between the two incoming streams, and $\Phi_{stag}$ is the image of the stagnation point where the two streams meet.
Such a solution is valid at high Peclet numbers, when the width of the diffusive wake (which scales as $\sqrt{\Pen \ \phi}$) is much smaller than the half-width of the channel. Here, we are modeling a very elongated domain, and must account for regions downstream where the mixing region becomes comparable to the channel width. In this case we can construct a solution in terms of Green's functions \citep{polyanin2001handbook}.
\begin{equation}
	c \left( \phi + \psi \ i \right) = a + \sum_{j=1}^{\infty} \frac{2}{\pi \ j \ \text{sin}\left(\pi \ j \ a \right)} \text{exp} \left( \left( \frac{\Pen}{2} - \sqrt{\pi^2 \ j^2 + \frac{\Pen^2}{4}} \right) \left( \phi - \phi_{stag} \right) \right) \text{cos} \left( \pi \ j \ \psi \right)
\end{equation}
Where $a$ is the dimensionless flow rate of the smaller of the two inlets (alternatively $a$ is the distance between the wake and the wall in the streamline coordinate problem, see \cite{boulais2025steady} for more details on the construction of solutions in the streamline coordinate domain.
This solution commits a small error around the stagnation point but becomes adequate further downstream when the diffusive wake's width becomes comparable to the width of the channel.
To minimize error, we use a combination of the wake solution in the near fiel and Helmholtz solution in the far-field to construct uniformly valid concentration profiles. Results are shown in figure \ref{fig_diffusion}.

\section{Limitations}

Before concluding this article, we highlight one or two limitations of the methods, as well as ways it could be extended to study more complex problems. 
One of the main limitation of our formulation, in terms of applications, is the neglect of no-slip boundary conditions at the walls of our planar microfluidic network. In many planar microfluidic systems, this neglect often does not lead to significant errors \citep{boulais20232d}, especially when the important physics happens near the center of the channels. However, when modeling porous media, or microfluidic models for disordered media, the presence of no-slip conditions at the walls may be a central factor in the flow physics at play. 
This can happen in a number of different ways. 
The most obvious one is the important contribution of no-slip boundaries to the flow resistance of the medium, as well as to high residence times of colloids passing through it, especially when channel depth becomes comparable to channel width. 
The second way in which potential flow formulation is limiting, even in very low Reynolds numbers flows, is in its inability to model vortices which may form near corners in Stokes flow \citep{dean1949steady, moffatt1964viscous}. Such vortices have recently been shown to play an important role in at least some families of flow through disordered media \citep{residori2025flow}.
In a similar vein, our method has trouble properly modeling flow in dead-end pores, especially very long ones. Dead-end pores are known to have important effects in applications related to porous media, in particular when studying bacterial transport or diffusiophoresis \citep{battat2019particle, de2021chemotaxis, alipour2026diffusiophoretic}.

\section{Possible Extensions}

The method presented here could be extended in a number of different directions.
The first and perhaps most obvious extension of our method is its application to other phenomena described by Laplace equation in complex microstructured media. For example, the solutions developed here could be directly applied to steady-state heat transport in thermal metamaterials \citep{schittny2013experiments}, or within certain limits to the study of mechanical metamaterials \citep{bertoldi2017flexible}.
%
Beyond purely Laplacian phenomena, certain other systems of partial differential equations can be studied using the same conformal mapping approach \citep{bazant2004conformal}. This is the case of the steady advection-diffusion system we showcased in section \ref{sec_diffusion}, but a similar method could also work for problems of ion transport in microfluidic systems \citep{gu2022electrokinetics}. 
The flow solutions described in this paper could also be used as a basis on which more complex numerical transport problems could be overlaid. For instance, particulate simulations could be combined with our flow solutions to model colloid transport in complex media \citep{zhang2012review}, or transport of active matter in the same media \citep{kurzthaler2021geometric}. Similarly, crystallization or other forms of agglomerative transport in microfluidic systems could be modeled by combining our solutions with either Eulerian population balance models \citep{ramkrishna2000population} or Lagrangian models of particle agglomeration \citep{meakin1983formation}.



\section{Conclusion}

In conclusion, we have shown a novel method for constructing flow solutions in complex geometries. By combining Schwarz-Christoffel maps, solutions for multipolar flows, and a segmentation method inspired by integrated circuit analysis, we have show how we can construct a library of base building blocks which can then be assembled at will to model elaborate microfluidic circuits. The numerical computations required are limited to the computation of the Schwarz-Christoffel map, which only has to be computed once for each building block geometry.
The models developed therein have direct applications to the study of microfluidic large-scale integration, as well as various problems of flow through porous media, and can be directly extended to a range of other Laplacian phenomena.
The models presented here are at the frontier of what can be done analytically (or semi-analytically) with ideal flow, and we hope they can serve as a starting point to stimulate further research in the field of disordered media and the physics of complex systems.

\appendix{Resistor Network Representation for Junction Element}\label{appendix_resistors}

\begin{figure}[h]
	\centering
	\includegraphics[width=0.95\textwidth]{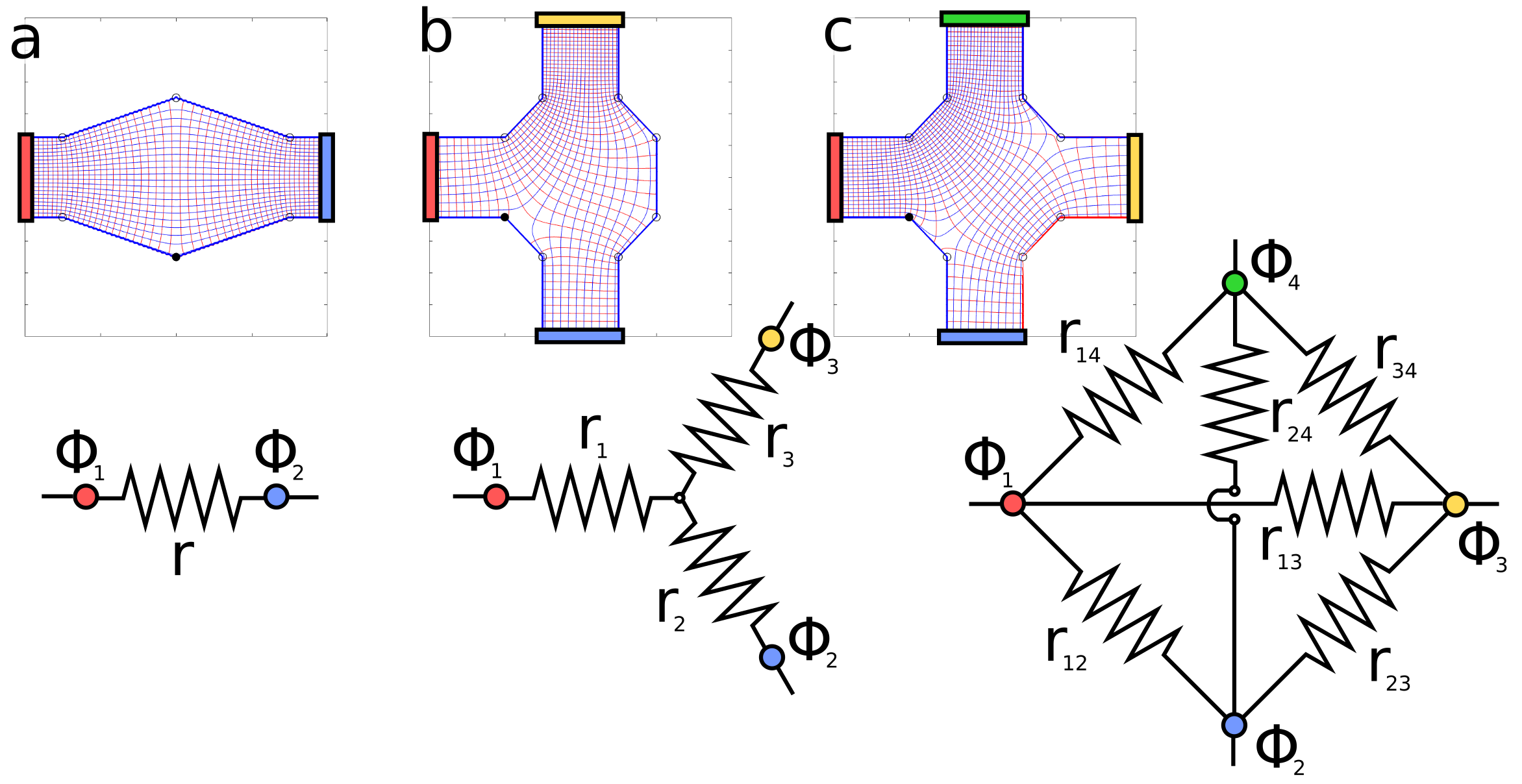}

	\caption{Representation of individual elements as resistor networks for linking potential and flow at the edges. a. A channel element can be represented by a single resistor. b. A 3-port junction can be represented using 3 resistors. c. A 4-port junction is represented as a network of 6 different resistors}
	\label{fig_resistors}
\end{figure}

The solution given in equation \ref{eq_sumoflogs}, combined with the appropriate Schwarz-Christoffel transform, gives us complete information about flow everywhere within a single junction element, provided we know the flow rates at its inlets and outlets. However, when assembling large numbers of elements in a complex network, the flow rate at every point in the network is not always known (flow rate and / or potential usually being prescribed at inlets and outlets of the entire circuit and not at the break lines between elements). In order to properly assemble our building blocks, we want to first know what the flow rates and potentials are at every break line, and from there the flow map can be built in each element using equation \ref{eq_sumoflogs}.

In order to determine the potential and flow rates at every junction, we use a representation of each element in terms of a network of resistor. The resistance of each element in this network can be determined by using equation \ref{eq_sumoflogs} and numerically measuring the potential difference between ports. Resistor network representation for 2, 3, and 4-port elements are shown in Fig. \ref{fig_resistors}.

The single resistance for a 2-port element can be determined by imposing a flow rate in the element, measuring the potential at the edges, and using the simple linear relation $\Delta \Phi = R Q$. Resistance in more complex elements require multiple measurements, imposing flow on every possible pair of ports. For elements with 4 or more ports, the network representation is no longer planar, and finding the individual value of resistors is better done using a numerical method.

Once the resistor network representation of a junction element has been determined, it can be integrated into complex circuits, and flow rates and potential at every break line can then be found using either Kirchoff's laws, freely available circuit solvers, or numerical relaxation methods in the case of more complex circuits (such as those showcased in section \ref{sec_porous}).



\bibliographystyle{jfm}
\bibliography{legosources}

\end{document}